\documentclass[10pt,conference]{IEEEtran}
\IEEEoverridecommandlockouts

\usepackage{cite}
\usepackage[T1]{fontenc}
\usepackage{graphicx}
\usepackage{amssymb}
\usepackage{amsmath}
\usepackage{amsthm}
\usepackage{paralist}
\usepackage[font=small]{caption}
\usepackage{subcaption}          
\usepackage{booktabs} 
\usepackage{algorithm}
\usepackage{multirow}
\usepackage{microtype} 
\usepackage{tablefootnote}
\usepackage{scrextend}
\usepackage{hyperref}
\usepackage{balance}
\usepackage{breqn}

\usepackage{amssymb}
\usepackage{amsfonts}
\usepackage{mathrsfs}
\usepackage{xspace}
\usepackage{bm}
\usepackage{upgreek}

\newcommand{\safemath}[2]{\newcommand{#1}{\ensuremath{#2}\xspace}}

\safemath{\bma}{\mathbf{a}}
\safemath{\bmb}{\mathbf{b}}
\safemath{\bmc}{\mathbf{c}}
\safemath{\bmd}{\mathbf{d}}
\safemath{\bme}{\mathbf{e}}
\safemath{\bmf}{\mathbf{f}}
\safemath{\bmg}{\mathbf{g}}
\safemath{\bmh}{\mathbf{h}}
\safemath{\bmi}{\mathbf{i}}
\safemath{\bmj}{\mathbf{j}}
\safemath{\bmk}{\mathbf{k}}
\safemath{\bml}{\mathbf{l}}
\safemath{\bmm}{\mathbf{m}}
\safemath{\bmn}{\mathbf{n}}
\safemath{\bmo}{\mathbf{o}}
\safemath{\bmp}{\mathbf{p}}
\safemath{\bmq}{\mathbf{q}}
\safemath{\bmr}{\mathbf{r}}
\safemath{\bms}{\mathbf{s}}
\safemath{\bmt}{\mathbf{t}}
\safemath{\bmu}{\mathbf{u}}
\safemath{\bmv}{\mathbf{v}}
\safemath{\bmw}{\mathbf{w}}
\safemath{\bmx}{\mathbf{x}}
\safemath{\bmy}{\mathbf{y}}
\safemath{\bmz}{\mathbf{z}}
\safemath{\bmzero}{\mathbf{0}}
\safemath{\bmone}{\mathbf{1}}

\bmdefine{\biad}{a}
\bmdefine{\bibd}{b}
\bmdefine{\bicd}{c}
\bmdefine{\bidd}{d}
\bmdefine{\bied}{e}
\bmdefine{\bifd}{f}
\bmdefine{\bigd}{g}
\bmdefine{\bihd}{h}
\bmdefine{\biid}{i}
\bmdefine{\bijd}{j}
\bmdefine{\bikd}{k}
\bmdefine{\bild}{l}
\bmdefine{\bimd}{m}
\bmdefine{\bind}{n}
\bmdefine{\biod}{o}
\bmdefine{\bipd}{p}
\bmdefine{\biqd}{q}
\bmdefine{\bird}{r}
\bmdefine{\bisd}{s}
\bmdefine{\bitd}{t}
\bmdefine{\biud}{u}
\bmdefine{\bivd}{v}
\bmdefine{\biwd}{w}
\bmdefine{\bixd}{x}
\bmdefine{\biyd}{y}
\bmdefine{\bizd}{z}

\bmdefine{\bixid}{\xi}
\bmdefine{\bilambdad}{\lambda}
\bmdefine{\bimud}{\mu}
\bmdefine{\bithetad}{\theta}
\bmdefine{\biphid}{\phi}
\bmdefine{\bideltad}{\delta}

\safemath{\bmia}{\biad}
\safemath{\bmib}{\bibd}
\safemath{\bmic}{\bicd}
\safemath{\bmid}{\bidd}
\safemath{\bmie}{\bied}
\safemath{\bmif}{\bifd}
\safemath{\bmig}{\bigd}
\safemath{\bmih}{\bihd}
\safemath{\bmii}{\biid}
\safemath{\bmij}{\bijd}
\safemath{\bmik}{\bikd}
\safemath{\bmil}{\bild}
\safemath{\bmim}{\bimd}
\safemath{\bmin}{\bind}
\safemath{\bmio}{\biod}
\safemath{\bmip}{\bipd}
\safemath{\bmiq}{\biqd}
\safemath{\bmir}{\bird}
\safemath{\bmis}{\bisd}
\safemath{\bmit}{\bitd}
\safemath{\bmiu}{\biud}
\safemath{\bmiv}{\bivd}
\safemath{\bmiw}{\biwd}
\safemath{\bmix}{\bixd}
\safemath{\bmiy}{\biyd}
\safemath{\bmiz}{\bizd}

\safemath{\bmxi}{\bixid}
\safemath{\bmlambda}{\bilambdad}
\safemath{\bmmu}{\bimud}
\safemath{\bmtheta}{\bithetad}
\safemath{\bmphi}{\biphid}
\safemath{\bmdelta}{\bideltad}

\safemath{\bA}{\mathbf{A}}
\safemath{\bB}{\mathbf{B}}
\safemath{\bC}{\mathbf{C}}
\safemath{\bD}{\mathbf{D}}
\safemath{\bE}{\mathbf{E}}
\safemath{\bF}{\mathbf{F}}
\safemath{\bG}{\mathbf{G}}
\safemath{\bH}{\mathbf{H}}
\safemath{\bI}{\mathbf{I}}
\safemath{\bJ}{\mathbf{J}}
\safemath{\bK}{\mathbf{K}}
\safemath{\bL}{\mathbf{L}}
\safemath{\bM}{\mathbf{M}}
\safemath{\bN}{\mathbf{N}}
\safemath{\bO}{\mathbf{O}}
\safemath{\bP}{\mathbf{P}}
\safemath{\bQ}{\mathbf{Q}}
\safemath{\bR}{\mathbf{R}}
\safemath{\bS}{\mathbf{S}}
\safemath{\bT}{\mathbf{T}}
\safemath{\bU}{\mathbf{U}}
\safemath{\bV}{\mathbf{V}}
\safemath{\bW}{\mathbf{W}}
\safemath{\bX}{\mathbf{X}}
\safemath{\bY}{\mathbf{Y}}
\safemath{\bZ}{\mathbf{Z}}

\safemath{\bZero}{\mathbf{0}}
\safemath{\bOne}{\mathbf{1}}
\safemath{\bDelta}{\mathbf{\Delta}}
\safemath{\bLambda}{\mathbf{\UpLambda}}
\safemath{\bPhi}{\mathbf{\Upphi}}
\safemath{\bSigma}{\mathbf{\Upsigma}}
\safemath{\bOmega}{\mathbf{\Upomega}}
\safemath{\bTheta}{\mathbf{\Uptheta}}

\bmdefine{\biAd}{A}
\bmdefine{\biBd}{B}
\bmdefine{\biCd}{C}
\bmdefine{\biDd}{D}
\bmdefine{\biEd}{E}
\bmdefine{\biFd}{F}
\bmdefine{\biGd}{G}
\bmdefine{\biHd}{H}
\bmdefine{\biId}{I}
\bmdefine{\biJd}{J}
\bmdefine{\biKd}{K}
\bmdefine{\biLd}{L}
\bmdefine{\biMd}{M}
\bmdefine{\biOd}{N}
\bmdefine{\biPd}{O}
\bmdefine{\biQd}{P}
\bmdefine{\biRd}{R}
\bmdefine{\biSd}{S}
\bmdefine{\biTd}{T}
\bmdefine{\biUd}{U}
\bmdefine{\biVd}{V}
\bmdefine{\biWd}{W}
\bmdefine{\biXd}{X}
\bmdefine{\biYd}{Y}
\bmdefine{\biZd}{Z}

\bmdefine{\biDelta}{\Delta}
\bmdefine{\biLambda}{\Lambda}
\bmdefine{\biPhi}{\Phi}
\bmdefine{\biSigma}{\Sigma}
\bmdefine{\biOmega}{\Omega}
\bmdefine{\biTheta}{\Theta}

\safemath{\bimA}{\biAd}
\safemath{\bimB}{\biBd}
\safemath{\bimC}{\biCd}
\safemath{\bimD}{\biDd}
\safemath{\bimE}{\biEd}
\safemath{\bimF}{\biFd}
\safemath{\bimG}{\biGd}
\safemath{\bimH}{\biHd}
\safemath{\bimI}{\biId}
\safemath{\bimJ}{\biJd}
\safemath{\bimK}{\biKd}
\safemath{\bimL}{\biLd}
\safemath{\bimM}{\biMd}
\safemath{\bimN}{\biNd}
\safemath{\bimO}{\biOd}
\safemath{\bimP}{\biPd}
\safemath{\bimQ}{\biQd}
\safemath{\bimR}{\biRd}
\safemath{\bimS}{\biSd}
\safemath{\bimT}{\biTd}
\safemath{\bimU}{\biUd}
\safemath{\bimV}{\biVd}
\safemath{\bimW}{\biWd}
\safemath{\bimX}{\biXd}
\safemath{\bimY}{\biYd}
\safemath{\bimZ}{\biZd}

\safemath{\bimDelta}{\biDelta}
\safemath{\bimLambda}{\biLambda}
\safemath{\bimPhi}{\biPhi}
\safemath{\bimSigma}{\biSigma}
\safemath{\bimOmega}{\biOmega}
\safemath{\bimTheta}{\biTheta}

\safemath{\setA}{\mathcal{A}}
\safemath{\setB}{\mathcal{B}}
\safemath{\setC}{\mathcal{C}}
\safemath{\setD}{\mathcal{D}}
\safemath{\setE}{\mathcal{E}}
\safemath{\setF}{\mathcal{F}}
\safemath{\setG}{\mathcal{G}}
\safemath{\setH}{\mathcal{H}}
\safemath{\setI}{\mathcal{I}}
\safemath{\setJ}{\mathcal{J}}
\safemath{\setK}{\mathcal{K}}
\safemath{\setL}{\mathcal{L}}
\safemath{\setM}{\mathcal{M}}
\safemath{\setN}{\mathcal{N}}
\safemath{\setO}{\mathcal{O}}
\safemath{\setP}{\mathcal{P}}
\safemath{\setQ}{\mathcal{Q}}
\safemath{\setR}{\mathcal{R}}
\safemath{\setS}{\mathcal{S}}
\safemath{\setT}{\mathcal{T}}
\safemath{\setU}{\mathcal{U}}
\safemath{\setV}{\mathcal{V}}
\safemath{\setW}{\mathcal{W}}
\safemath{\setX}{\mathcal{X}}
\safemath{\setY}{\mathcal{Y}}
\safemath{\setZ}{\mathcal{Z}}
\safemath{\emptySet}{\varnothing}

\safemath{\colA}{\mathscr{A}}
\safemath{\colB}{\mathscr{B}}
\safemath{\colC}{\mathscr{C}}
\safemath{\colD}{\mathscr{D}}
\safemath{\colE}{\mathscr{E}}
\safemath{\colF}{\mathscr{F}}
\safemath{\colG}{\mathscr{G}}
\safemath{\colH}{\mathscr{H}}
\safemath{\colI}{\mathscr{I}}
\safemath{\colJ}{\mathscr{J}}
\safemath{\colK}{\mathscr{K}}
\safemath{\colL}{\mathscr{L}}
\safemath{\colM}{\mathscr{M}}
\safemath{\colN}{\mathscr{N}}
\safemath{\colO}{\mathscr{O}}
\safemath{\colP}{\mathscr{P}}
\safemath{\colQ}{\mathscr{Q}}
\safemath{\colR}{\mathscr{R}}
\safemath{\colS}{\mathscr{S}}
\safemath{\colT}{\mathscr{T}}
\safemath{\colU}{\mathscr{U}}
\safemath{\colV}{\mathscr{V}}
\safemath{\colW}{\mathscr{W}}
\safemath{\colX}{\mathscr{X}}
\safemath{\colY}{\mathscr{Y}}
\safemath{\colZ}{\mathscr{Z}}

\safemath{\opA}{\mathbb{A}}
\safemath{\opB}{\mathbb{B}}
\safemath{\opC}{\mathbb{C}}
\safemath{\opD}{\mathbb{D}}
\safemath{\opE}{\mathbb{E}}
\safemath{\opF}{\mathbb{F}}
\safemath{\opG}{\mathbb{G}}
\safemath{\opH}{\mathbb{H}}
\safemath{\opI}{\mathbb{I}}
\safemath{\opJ}{\mathbb{J}}
\safemath{\opK}{\mathbb{K}}
\safemath{\opL}{\mathbb{L}}
\safemath{\opM}{\mathbb{M}}
\safemath{\opN}{\mathbb{N}}
\safemath{\opO}{\mathbb{O}}
\safemath{\opP}{\mathbb{P}}
\safemath{\opQ}{\mathbb{Q}}
\safemath{\opR}{\mathbb{R}}
\safemath{\opS}{\mathbb{S}}
\safemath{\opT}{\mathbb{T}}
\safemath{\opU}{\mathbb{U}}
\safemath{\opV}{\mathbb{V}}
\safemath{\opW}{\mathbb{W}}
\safemath{\opX}{\mathbb{X}}
\safemath{\opY}{\mathbb{Y}}
\safemath{\opZ}{\mathbb{Z}}
\safemath{\opZero}{\mathbb{O}}
\safemath{\identityop}{\opI}

\safemath{\veca}{\bma}
\safemath{\vecb}{\bmb}
\safemath{\vecc}{\bmc}
\safemath{\vecd}{\bmd}
\safemath{\vece}{\bme}
\safemath{\vecf}{\bmf}
\safemath{\vecg}{\bmg}
\safemath{\vech}{\bmh}
\safemath{\veci}{\bmi}
\safemath{\vecj}{\bmj}
\safemath{\veck}{\bmk}
\safemath{\vecl}{\bml}
\safemath{\vecm}{\bmm}
\safemath{\vecn}{\bmn}
\safemath{\veco}{\bmo}
\safemath{\vecp}{\bmp}
\safemath{\vecq}{\bmq}
\safemath{\vecr}{\bmr}
\safemath{\vecs}{\bms}
\safemath{\vect}{\bmt}
\safemath{\vecu}{\bmu}
\safemath{\vecv}{\bmv}
\safemath{\vecw}{\bmw}
\safemath{\vecx}{\bmx}
\safemath{\vecy}{\bmy}
\safemath{\vecz}{\bmz}

\safemath{\veczero}{\bmzero}
\safemath{\vecone}{\bmone}
\safemath{\vecxi}{\bmxi}
\safemath{\veclambda}{\bmlambda}
\safemath{\vecmu}{\bmmu}
\safemath{\vectheta}{\bmtheta}
\safemath{\vecphi}{\bmphi}
\safemath{\vecdelta}{\bmdelta}

\safemath{\matA}{\bA}
\safemath{\matB}{\bB}
\safemath{\matC}{\bC}
\safemath{\matD}{\bD}
\safemath{\matE}{\bE}
\safemath{\matF}{\bF}
\safemath{\matG}{\bG}
\safemath{\matH}{\bH}
\safemath{\matI}{\bI}
\safemath{\matJ}{\bJ}
\safemath{\matK}{\bK}
\safemath{\matL}{\bL}
\safemath{\matM}{\bM}
\safemath{\matN}{\bN}
\safemath{\matO}{\bO}
\safemath{\matP}{\bP}
\safemath{\matQ}{\bQ}
\safemath{\matR}{\bR}
\safemath{\matS}{\bS}
\safemath{\matT}{\bT}
\safemath{\matU}{\bU}
\safemath{\matV}{\bV}
\safemath{\matW}{\bW}
\safemath{\matX}{\bX}
\safemath{\matY}{\bY}
\safemath{\matZ}{\bZ}
\safemath{\matzero}{\bmzero}

\safemath{\matDelta}{\bDelta}
\safemath{\matLambda}{\bLambda}
\safemath{\matPhi}{\bPhi}
\safemath{\matSigma}{\bSigma}
\safemath{\matOmega}{\bOmega}
\safemath{\matTheta}{\bTheta}

\safemath{\matidentity}{\matI}
\safemath{\matone}{\matO}

\safemath{\rnda}{A}
\safemath{\rndb}{B}
\safemath{\rndc}{C}
\safemath{\rndd}{D}
\safemath{\rnde}{E}
\safemath{\rndf}{F}
\safemath{\rndg}{G}
\safemath{\rndh}{H}
\safemath{\rndi}{I}
\safemath{\rndj}{J}
\safemath{\rndk}{K}
\safemath{\rndl}{L}
\safemath{\rndm}{M}
\safemath{\rndn}{N}
\safemath{\rndo}{O}
\safemath{\rndp}{P}
\safemath{\rndq}{Q}
\safemath{\rndr}{R}
\safemath{\rnds}{S}
\safemath{\rndt}{T}
\safemath{\rndu}{U}
\safemath{\rndv}{V}
\safemath{\rndw}{W}
\safemath{\rndx}{X}
\safemath{\rndy}{Y}
\safemath{\rndz}{Z}

\safemath{\rveca}{\bimA}
\safemath{\rvecb}{\bimB}
\safemath{\rvecc}{\bimC}
\safemath{\rvecd}{\bimD}
\safemath{\rvece}{\bimE}
\safemath{\rvecf}{\bimF}
\safemath{\rvecg}{\bimG}
\safemath{\rvech}{\bimH}
\safemath{\rveci}{\bimI}
\safemath{\rvecj}{\bimJ}
\safemath{\rveck}{\bimK}
\safemath{\rvecl}{\bimL}
\safemath{\rvecm}{\bimM}
\safemath{\rvecn}{\bimN}
\safemath{\rveco}{\bomO}
\safemath{\rvecp}{\bimP}
\safemath{\rvecq}{\bimQ}
\safemath{\rvecr}{\bimR}
\safemath{\rvecs}{\bimS}
\safemath{\rvect}{\bimT}
\safemath{\rvecu}{\bimU}
\safemath{\rvecv}{\bimV}
\safemath{\rvecw}{\bimW}
\safemath{\rvecx}{\bimX}
\safemath{\rvecy}{\bimY}
\safemath{\rvecz}{\bimZ}

\safemath{\rvecxi}{\bmxi}
\safemath{\rveclambda}{\bmlambda}
\safemath{\rvecmu}{\bmmu}
\safemath{\rvectheta}{\bmtheta}
\safemath{\rvecphi}{\bmphi}

\safemath{\rmatA}{\bimA}
\safemath{\rmatB}{\bimB}
\safemath{\rmatC}{\bimC}
\safemath{\rmatD}{\bimD}
\safemath{\rmatE}{\bimE}
\safemath{\rmatF}{\bimF}
\safemath{\rmatG}{\bimG}
\safemath{\rmatH}{\bimH}
\safemath{\rmatI}{\bimI}
\safemath{\rmatJ}{\bimJ}
\safemath{\rmatK}{\bimK}
\safemath{\rmatL}{\bimL}
\safemath{\rmatM}{\bimM}
\safemath{\rmatN}{\bimN}
\safemath{\rmatO}{\bimO}
\safemath{\rmatP}{\bimP}
\safemath{\rmatQ}{\bimQ}
\safemath{\rmatR}{\bimR}
\safemath{\rmatS}{\bimS}
\safemath{\rmatT}{\bimT}
\safemath{\rmatU}{\bimU}
\safemath{\rmatV}{\bimV}
\safemath{\rmatW}{\bimW}
\safemath{\rmatX}{\bimX}
\safemath{\rmatY}{\bimY}
\safemath{\rmatZ}{\bimZ}

\safemath{\rmatDelta}{\bimDelta}
\safemath{\rmatLambda}{\bimLambda}
\safemath{\rmatPhi}{\bimPhi}
\safemath{\rmatSigma}{\bimSigma}
\safemath{\rmatOmega}{\bimOmega}
\safemath{\rmatTheta}{\bimTheta}
\usepackage{amssymb}
\usepackage{amsfonts}
\usepackage{mathrsfs}
\usepackage{xspace}
\usepackage{bm}
\usepackage{fancyref}
\usepackage{textcomp}

\usepackage{multirow}
\usepackage{stmaryrd}

\newenvironment{textbmatrix}{	\setlength{\arraycolsep}{2.5pt}%
	\big[\begin{matrix}}{\end{matrix}\big]%
	\raisebox{0.08ex}{\vphantom{M}}}

\def\be{\begin{equation}}
	\def\ee{\end{equation}}
\def\een{\nonumber \end{equation}}
\def\mat{\begin{bmatrix}}
\def\emat{\end{bmatrix}}
\def\btm{\begin{textbmatrix}}
\def\etm{\end{textbmatrix}}

\def\ba#1\ea{\begin{align}#1\end{align}}
\def\bas#1\eas{\begin{align*}#1\end{align*}}
\def\bs#1\es{\begin{split}#1\end{split}}
\def\bg#1\eg{\begin{gather}#1\end{gather}}
\def\bml#1\eml{\begin{multline}#1\end{multline}}
\def\bi#1\ei{\begin{itemize}#1\end{itemize}}

\newcommand{\lefto}{\mathopen{}\left}

\DeclareMathOperator{\Exop}{\opE}			

\newcommand{\Ex}[2]{\ensuremath{\Exop_{#1}\lefto[#2\right]}} 	

\safemath{\dirac}{\delta}					
\safemath{\krond}{\dirac}					

\safemath{\upto}{\uparrow}
\safemath{\downto}{\downarrow}
\safemath{\iu}{j}							
\safemath{\ev}{\lambda}						
\safemath{\hilseqspace}{l^{2}}				
\newcommand{\banachfunspace}[1]{\setL^{#1}}	
\safemath{\hilfunspace}{\banachfunspace{2}}	

\safemath{\SNR}{\textit{SNR}} 				
\safemath{\PAR}{\textit{PAR}} 				
\safemath{\No}{N_0}							
\safemath{\Es}{E_s}							
\safemath{\Eb}{E_b}							
\safemath{\EbNo}{\frac{\Eb}{\No}}
\safemath{\EsNo}{\frac{\Es}{\No}}

\DeclareMathOperator{\CHop}{\ensuremath{\opH}} 
\safemath{\tvir}{\rndh_{\CHop}}				
\safemath{\tvtf}{\rndl_{\CHop}}				
\safemath{\spf}{\rnds_{\CHop}}				
\safemath{\bff}{H_{\CHop}}					

\safemath{\ircf}{r_{h}}						
\safemath{\tftvcf}{r_{s}}					
\safemath{\tfcf}{r_{l}}						
\safemath{\bfcf}{r_{H}}						

\safemath{\tcorr}{c_h}						
\safemath{\scf}{c_{s}}						
\safemath{\tfcorr}{c_{l}}					
\safemath{\fcorr}{c_{H}}						

\safemath{\mi}{I}							
\safemath{\capacity}{C}						

\safemath{\normal}{\mathcal{N}}			
\safemath{\jpg}{\mathcal{CN}}			
\safemath{\mchain}{\leftrightarrow}		

\safemath{\dB}{\,\mathrm{dB}}
\safemath{\dBm}{\,\mathrm{dBm}}
\safemath{\Hz}{\,\mathrm{Hz}}
\safemath{\kHz}{\,\mathrm{kHz}}
\safemath{\MHz}{\,\mathrm{MHz}}
\safemath{\GHz}{\,\mathrm{GHz}}
\safemath{\s}{\,\mathrm{s}}
\safemath{\ms}{\,\mathrm{ms}}
\safemath{\mus}{\,\mathrm{\text{\textmu}s}}
\safemath{\ns}{\,\mathrm{ns}}
\safemath{\ps}{\,\mathrm{ps}}
\safemath{\meter}{\,\mathrm{m}}
\safemath{\mm}{\,\mathrm{mm}}
\safemath{\cm}{\,\mathrm{cm}}
\safemath{\m}{\,\mathrm{m}}
\safemath{\W}{\,\mathrm{W}}
\safemath{\mW}{\, \mathrm{mW}}
\safemath{\J}{\,\mathrm{J}}
\safemath{\K}{\,\mathrm{K}}
\safemath{\bit}{\,\mathrm{bit}}
\safemath{\nat}{\,\mathrm{nat}}

\safemath{\define}{\triangleq}			

\safemath{\equivalent}{\sim}
\safemath{\distas}{\sim}					
\safemath{\sdiff}{\Delta}				

\safemath{\reals}{\mathbb{R}}
\safemath{\positivereals}{\reals_{+}}
\safemath{\integers}{\mathbb{Z}}
\safemath{\posint}{\integers_{+}}
\safemath{\naturals}{\mathbb{N}}
\safemath{\posnaturals}{\naturals_{+}}
\safemath{\complexset}{\mathbb{C}}
\safemath{\rationals}{\mathbb{Q}}

\newcommand*{\fancyrefapplabelprefix}{app}		
\newcommand*{\fancyrefthmlabelprefix}{thm}		
\newcommand*{\fancyreflemlabelprefix}{lem}		
\newcommand*{\fancyrefcorlabelprefix}{cor}		
\newcommand*{\fancyrefdeflabelprefix}{def}		
\newcommand*{\fancyrefproplabelprefix}{prop}		
\newcommand*{\fancyrefexmpllabelprefix}{exmpl}
\newcommand*{\fancyrefalglabelprefix}{alg}		
\newcommand*{\fancyreftbllabelprefix}{tbl}		

\frefformat{vario}{\fancyrefseclabelprefix}{Section~#1}
\frefformat{vario}{\fancyrefthmlabelprefix}{Theorem~#1}
\frefformat{vario}{\fancyreftbllabelprefix}{Table~#1}
\frefformat{vario}{\fancyreflemlabelprefix}{Lemma~#1}
\frefformat{vario}{\fancyrefcorlabelprefix}{Corollary~#1}
\frefformat{vario}{\fancyrefdeflabelprefix}{Definition~#1}
\frefformat{vario}{\fancyreffiglabelprefix}{Figure~#1}
\frefformat{vario}{\fancyrefapplabelprefix}{Appendix~#1}
\frefformat{vario}{\fancyrefeqlabelprefix}{(#1)}
\frefformat{vario}{\fancyrefproplabelprefix}{Proposition~#1}
\frefformat{vario}{\fancyrefexmpllabelprefix}{Example~#1}
\frefformat{vario}{\fancyrefalglabelprefix}{Algorithm~#1}

\usepackage{color}

\usepackage{enumitem}
\setlist[itemize]{leftmargin=*, itemsep=0.3em, topsep=0.3em} 
\allowdisplaybreaks 

\DeclareRobustCommand{\relief}{
	\includegraphics[width=0.3cm]{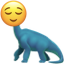}%
}

\begin{document}


\title{Variable Point: A Number Format for Area- and Energy-Efficient Multiplication of High-Dynamic-Range Numbers}
\author{
	\IEEEauthorblockN{Seyed Hadi Mirfarshbafan, Nicolas Filliol, Oscar Castañeda, and Christoph Studer}\\[-0.2cm]
	\IEEEauthorblockA{\em Department of Information Technology and Electrical Engineering, ETH Zurich, Switzerland\\
		email: mirfarshbafan@iis.ee.ethz.ch, nfilliol@student.ethz.ch, caoscar@ethz.ch, and studer@ethz.ch}
		\thanks{This paper summarizes Chapter 6 of the doctoral thesis \cite{seyedhadi_thesis}.}
		\thanks{This work was funded in part by the Swiss State Secretariat for Education, Research, and Innovation (SERI) under the SwissChips initiative. \relief{}}
}

\maketitle

\begin{abstract}

Fixed-point number representation is commonly employed in digital VLSI designs that have stringent hardware efficiency constraints. However, fixed-point numbers cover a relatively small dynamic range for a given bitwidth. In contrast, floating-point numbers offer a larger dynamic range at the cost of increased hardware complexity. In this paper, we propose a novel number format called variable-point (VP). VP numbers cover a larger dynamic range than fixed-point numbers with similar bitwidth, without notably increasing hardware complexity---this allows for a more efficient representation of signals with high dynamic range. To demonstrate the efficacy of the proposed VP number format, we consider a matrix-vector multiplication engine for spatial equalization in multi-antenna wireless communication systems involving high-dynamic-range signals. Through post-layout VLSI implementation results, we demonstrate that the proposed VP-based design achieves 20\% and 10\% area and power savings, respectively, compared to a fully optimized fixed-point design, without incurring any noticeable performance degradation.

\end{abstract}
	
\section{Introduction}

In digital hardware, the number representation format determines quantization errors as well as the resulting hardware efficiency. Among the prominent formats, fixed-point (FXP) offers the best hardware efficiency, due to simple arithmetic components, but covers a relatively small dynamic range.
Another prominent number format is floating-point (FLP), which is used in applications with high-dynamic-range signals, including in the training stage of deep neural networks~\cite{nips17, courbariaux2015}, as well as in general purpose hardware. The main drawback of the floating-point format is the high complexity of floating-point arithmetic hardware. Therefore, alternative number formats have been proposed for energy- and resource-constrained systems, with the goal of combining the efficiency of FXP with the high-dynamic-range support of FLP. 

A prominent hybrid number format is block floating-point (BFP), which improves implementation efficiency by sharing a single exponent among a block of numbers, each with a separate mantissa~\cite{Wilkinson94}. 
A common approach to determining the shared exponent is setting it to the largest exponent among the individual FLP representations of the block elements. 
BFP offers a trade-off between accuracy and complexity controlled by the block size, the shared exponent choice, as well as the bitwidth of the mantissas and the shared exponent. 
Although BFP strikes a balance between the efficiency of FXP and the flexibility and high-dynamic-range coverage of FLP, it is, in general, still less hardware efficient than a FXP implementation with similar significand bitwidths.

\subsubsection*{Contributions}

In this paper, we propose a novel number format, called variable-point (VP), which provides a larger dynamic range than the FXP format for the same significand bitwidth, without notable hardware overhead. 
This allows the use of lower-resolution VP significands in applications involving high-dynamic-range signals, thereby reducing area and power compared to an FXP implementation.
We showcase the efficacy of VP numbers in a target application with high-dynamic-range signals, through post-layout very large-scale integration (VLSI) implementation results, and demonstrate that utilizing VP can lead to circuits with lower area and power compared to a fully-optimized FXP implementation.

\section{Variable-Point (VP) Numbers} \label{sec:VP}

\begin{figure}[t]
	\centering
	\includegraphics[width=0.85\linewidth]{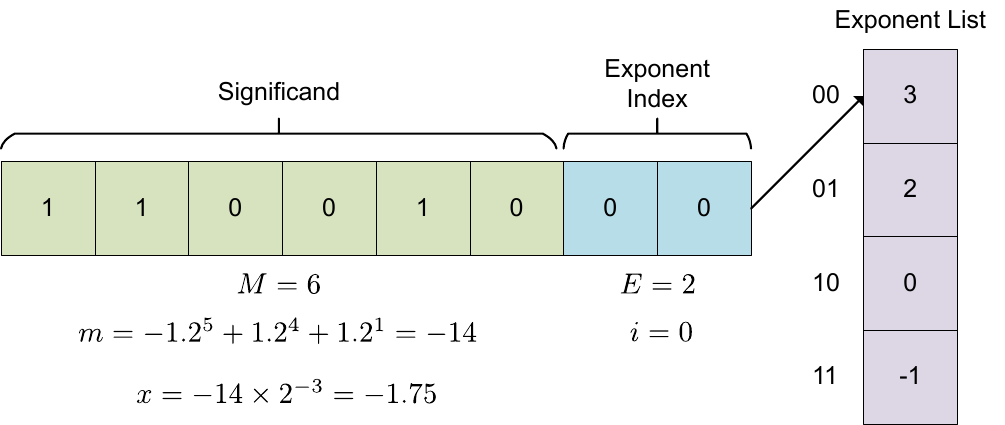}
	\caption{An example of a VP number $x$ with $M=6$ significand bits and $E=2$ exponent bits and the exponent list $\bmf=[3,2,0,-1]$.}
	\label{fig:example}
\end{figure}

The proposed VP number format consists of two fields: (i)~an $M$-bit significand $m$, which is a two's complement integer, and (ii)~an $E$-bit \emph{exponent index} $i$, which points to one of the $2^E$ exponent options. Implicit in any VP representation is a vector of $2^E$ exponent options, referred to as the \emph{exponent list}~$\bmf$. The real number represented by this format is given by 
\begin{align}
	x=m \times 2^{-f_i},
\end{align}
where $f_i$ is the $i$th entry of the exponent list. An example is given in \fref{fig:example}. In fact, $\bmf$ is the list of possible fractional lengths, hence, we multiply $m$ with $2^{-f_i}$ rather than $2^{f_i}$.

We use $\text{VP}(M, \bmf)$ to designate a VP number with an $M$-bit significand and the exponent list $\bmf$. The number of bits for exponent index is implicitly given by $E = \log_2(|\bmf|)$, where we slightly abuse the cardinality notation $|\bmf|$ to denote the dimension of the vector $\bmf$. Note that throughout the paper, we assume that $|\bmf|$ is a power of $2$. Furthermore, we will use the notation $\text{FXP}(W, F)$ to denote a $W$-bit two's complement FXP number with $F$ fractional bits.

The idea behind the VP format is to represent a high-resolution FXP number with a lower-resolution significand, by selecting the most important bit range of the original high-resolution FXP number. We select the bit range such that the precision loss is minimized, while not incurring any overflow, as discussed in \fref{sec:FXP2VP}.

\subsection{VP Arithmetic} \label{sec:arith}

The proposed VP-based implementation scheme is based on converting high-resolution FXP numbers to VP numbers with a lower-resolution significand.
By reducing the resolution of the original FXP number, the VP-based design allows for reduction in the area of arithmetic components.

As will be detailed shortly, a VP multiplier is simply a FXP multiplier whose operands are the significands of the VP inputs. 
While it is possible to build adders/subtractors with VP inputs, in this paper, we use VP exclusively for multiplication and perform all additions and subtractions using FXP numbers. To this end, the inputs to the multipliers are converted from FXP to VP as detailed in \fref{sec:FXP2VP}, and the result is converted back to FXP for further processing, as detailed in \fref{sec:VP2FXP}, unless the next operation is another multiplication. In order to achieve the goal of reducing the area of multipliers, the savings achieved by the reduction in significand bitwidths through VP conversion needs to outweigh the overhead due to the conversions. We will investigate this trade-off in \fref{sec:impl}.

\subsection{Multiplication with VP Inputs}

Consider two VP numbers $a$ and $b$, with formats VP$(M_a, \bmf_a)$ and VP$(M_b, \bmf_b)$, respectively. Let $m_a$ and $m_b$ denote the significands and $i_a$ and $i_b$ denote the exponent indices of~$a$ and $b$, respectively. As with standard FLP, the multiplication of two VP numbers amounts to multiplying the significands to obtain the significand of the product, and adding the exponents to obtain the exponent of the product. However, one of the main advantages of VP over FLP is that there is no need to actually add the exponents to obtain the product exponent. In fact, recall that the VP numbers do not explicitly carry the exponent lists $\bmf_a$ and $\bmf_b$, but rather the exponent indices. The actual exponent lists are provided as parameters to the conversion units. In particular, the exponent list of the product is constructed offline as the pairwise sum of the entries from $\bmf_a$ and $\bmf_b$, and is provided as a parameter to the VP-to-FXP converter.
Therefore, to obtain the exponent list of the product, the multiplier simply concatenates the exponent indices of inputs.

\subsection{FXP-to-VP (FXP2VP) Conversion} \label{sec:FXP2VP}

VP numbers are envisioned to coexist with FXP numbers in the same hardware. Hence, it is crucial to have a hardware-friendly approach for conversion between FXP and VP. Consider a FXP$(W, F)$ number $x_{\text{FXP}}$ which we want to convert into $x_{\text{VP}}$ with VP$(M, \bmf)$ format. 
For the conversion from FXP to VP to be useful, we assume that $W > M$ and $F \geq \max(\bmf)$. The conversion, however, is still possible without these assumptions, but requires sign extension and zero padding. The conversion consists of the following two steps:

\begin{itemize}
	\item \emph{Set the exponent index $i$:} Identify the index $i$ of the largest entry of $\bmf$ such that the integer part of $x_{\text{FXP}}$ fits into the remaining $M - f_i$ bits without overflow. Set the exponent index of $x_{\text{VP}}$ to $i$.
	\item \emph{Set the significand $m$:} Select $f_i$ bits to the right of the binary point and $M-f_i$ bits to the left of the binary point from $x_{\text{FXP}}$ as the $M$-bit significand of $x_{\text{VP}}$.
\end{itemize}

\begin{figure}[t]
	\centering
	\includegraphics[width=0.9\linewidth]{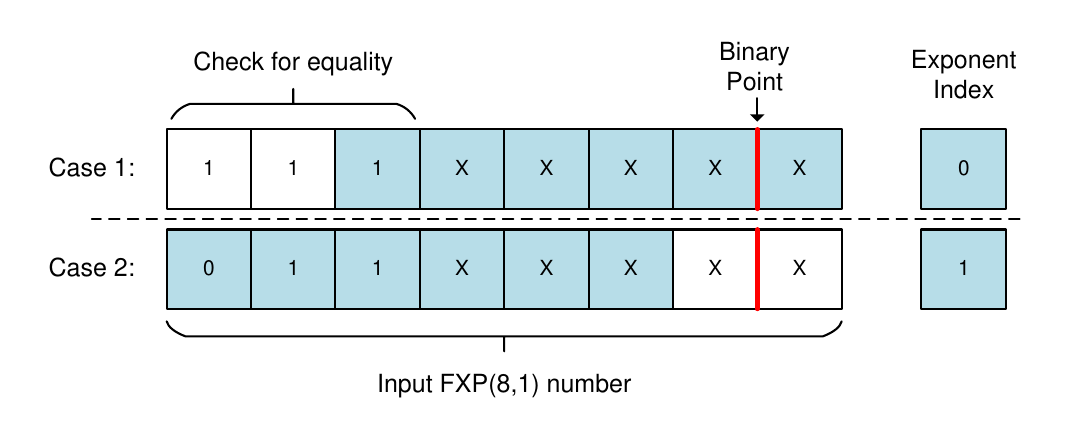}
	\caption{Two examples illustrating the conversion from FXP$(8,1)$ to VP$(6,[1,-1])$. The shaded bits show the significand of the converted VP number and the exponent index is shown separately.}
	\label{fig:FXP2VP_illustration}
\end{figure}

This procedure is illustrated in \fref{fig:FXP2VP_illustration} for two example inputs in FXP$(8,1)$ format, which are converted into VP$(6,[1,-1])$, in which case $E=1$. To determine $i$, we check the $W-M+1=3$ MSBs of the input FXP number. If these three bits are equal, then we set $i=0$ and select the lower $6$ bits of the input as the significand of the VP number; otherwise we set $i=1$ and select the upper $6$ bits.

\begin{figure}[t]
	\centering
	\includegraphics[width=0.7\linewidth]{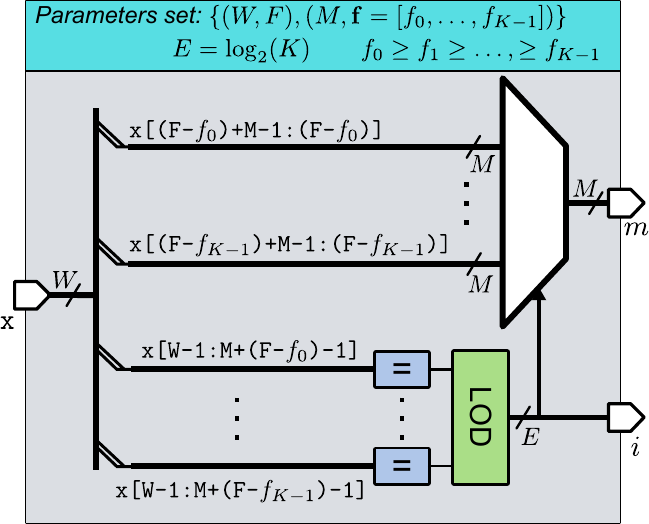}
	\caption{Architecture of FXP2VP converter parameterized by $\{(W, F), (M, \bmf)\}$. The converter takes the FXP input $x$ and produces the significand $m$ and the exponent index $i$ of the corresponding VP number.}
	\label{fig:FXP2VP}
\end{figure}

An efficient VLSI architecture for the FXP2VP conversion procedure described above is illustrated in \fref{fig:FXP2VP}. The purely combinational FXP2VP is parameterized with the set $\{(W, F), (M, \bmf)\}$; each instance is synthesized with the given parameters and the parameters cannot change once the circuit is synthesized. Note that for the proposed architecture to function properly, the entries of the exponent list $\bmf$ need to be sorted in descending order, i.e., $f_0 \geq f_1 \geq \ldots \geq f_{K-1}$, where $K=2^E$.
In this paper, we adopt the Verilog notation to denote the bit ranges of multi-bit signals. For a $W$-bit signal $\texttt{x}$, the MSB is $\texttt{x[W-1]}$, the LSB is $\texttt{x[0]}$ and $\texttt{x[W-1:1]}$ denotes all bits except the LSB.

The FXP2VP module takes a $W$-bit FXP input $\texttt{x}$ with $F$~fractional bits and operates as follows: for each fractional length option $f_k$, $k=0, \ldots, K-1$, it checks the MSBs specified by $\texttt{x[W-1:M+(F-}f_k\texttt{)-1]}$ for equality and passes the result to a leading-one detector (LOD) circuit, 
which determines the smallest $i$ (i.e., largest $f_i$) for which the corresponding MSBs are all $1$ or all $0$. The output of the LOD is the exponent index~$i$, which also selects $\texttt{x[(F-}f_i\texttt{)+M-1:(F-}f_i\texttt{)]}$ as the output significand.

\subsection{Parameter Selection}
Conversion from FXP to VP is useful only when the significand bitwidth of the VP number is smaller than the bitwidth of the input FXP number.  
Converting a FXP$(W, F)$ number into VP$(M, \bmf)$ format with $W > M$ would generally result in a precision loss.
The choice of the VP parameters ($M$ and $\bmf$) determine the trade-off between hardware efficiency and precision loss. The optimal parameters are determined for each signal individually using Monte-Carlo simulations to ensure that the precision loss is negligible for the target application. In general, we set $\max(\bmf) = F$, so that the VP format has sufficient resolution for input FXP numbers with small magnitude. Additionally, we set $\min(\bmf)$ such that $W-F = M - \min(\bmf)$, to ensure that the VP number has enough integer bits to accommodate all numbers without~overflow.

\subsection{VP-to-FXP (VP2FXP) Conversion} \label{sec:VP2FXP}

\begin{figure}[t]
	\centering
	\includegraphics[width=0.95\linewidth]{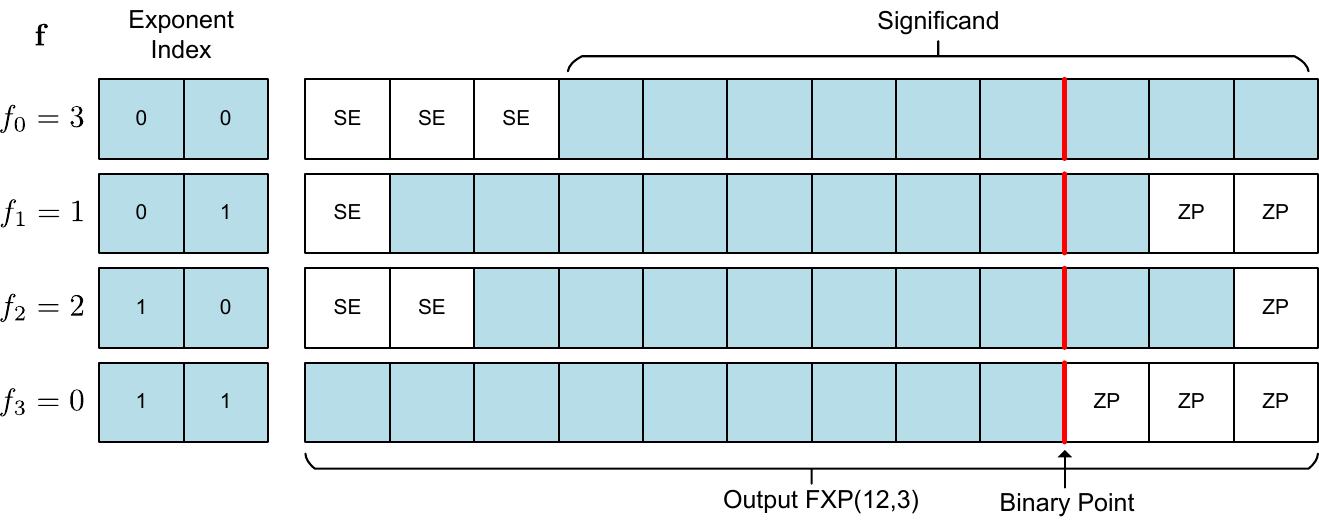}
	\caption{An example illustration of conversion from VP$(9,[3,1,2,0])$ to FXP$(12,3)$. For each of the four exponent options, we put the significand in the appropriate bit range of the output FXP number, and sign extend (SE) the remaining MSBs and zero-pad (ZP) the remaining LSBs.}
	\label{fig:VP2FXP_illustration}
\end{figure}

Consider converting a VP$(M, \bmf)$ number into a FXP$(W, F)$ number. 
The conversion amounts to zero padding and right shifting the $M$-bit significand, based on the exponent index of the input. The exposed MSBs of the output are filled by sign extension. \fref{fig:VP2FXP_illustration} illustrates an example.

\begin{figure}[t]
	\centering
	\includegraphics[width=0.7\linewidth]{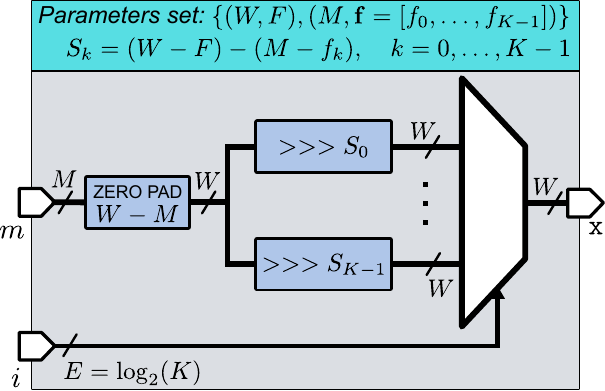}
	\caption{Architecture of VP2FXP converter parameterized by $\{(W, F), (M, \bmf)\}$. The converter takes the significand $m$ and the exponent index $i$ and produces the corresponding FXP number $\texttt{x}$.}
	\label{fig:VP2FXP}
\end{figure}

An efficient architecture for the VP2FXP conversion described is illustrated in \fref{fig:VP2FXP}.
Similar to the FXP2VP converter, VP2FXP is also purely combinational and parameterized by the set $\{(W, F), (M, \bmf)\}$; each instance is synthesized for a given set of parameters.
The $M$-bit significand input is first zero-padded with $W-M$ LSBs. The resulting $W$-bit number then gets arithmetically right shifted (i.e., the deserted MSBs are filled by sign extension) by $S_k = (W-F)- (M-f_k)$ bits for $k=0, \ldots, K-1$, where $K=2^E$. The exponent index of the input determines which shifted version of the zero-padded significand is selected as the FXP output $\texttt{x}$.

\subsection{Comparison to Floating-Point}

At first glance, VP looks similar to the FLP format. However, there are fundamental differences that make the VP-based implementations more efficient than not only FLP-based, but also FXP-based implementations with similar performance.  
The key differences of VP and FLP are summarized below:

\begin{itemize}
	
	\item \emph{Arbitrary exponent list:} In the standard FLP format, the exponents cover a contiguous range of integers, while in VP one can choose the exponent list (i.e., the fractional length options) arbitrarily, allowing for a more customized format for each signal, which maximizes the representation efficiency and simplifies arithmetic hardware.
	\item \emph{Arbitrary scale:} Somewhat related to the above point, the VP format can be tuned to fully utilize all the available bits to represent signals of arbitrary scale. To clarify this point, imagine that a particular signal in the design only takes integer values. Representing such a signal with FLP effectively wastes the negative exponents, as they are never used for such a signal. In other words, the fact that the exponent values form a continous range of integers centered around zero (after subtracting the bias), can result in underutilized exponent values for numbers with certain characteristics. In contrast, VP enables one to choose the exponent options arbitrarily. 
	\item \emph{Separate format for each signal:} Floating-point arithmetic units typically have the same format for inputs and outputs, which generally results in a unified format throughout the design. 
	This can be suboptimal as all signals must be expressed with a format that satisfies the worst-case signal. 
	In contrast, in designs using VP, each VP signal has its own optimized parameters, allowing for maximum efficiency.
	\item  \emph{More compatible with FXP:} Since the VP format uses two's complement 
	FXP for the significand, conversion from and to FXP is simple and efficient. Furthermore, this enables one to implement VP multiplication by simply multiplying the significands of the operands using standard FXP multipliers.
	\item \emph{Efficient multiplication:} As noted in \fref{sec:arith}, in a VP multiplier, there is no need to add the exponents of the operands, as opposed to FLP.
	
\end{itemize} 

Finally, we acknowledge that VP is a customized number format which is suitable for application-specific hardware implementations, and does not provide the versatility and universality of the standard FLP format. Furthermore, VP is efficient for small $E$ and its efficiency compared to FXP degrades for large values of $E$, as conversion becomes costly.

\section{Example Application} \label{sec:beamspace}

\begin{figure}[t]
	\centering
	\includegraphics[width=0.9\linewidth]{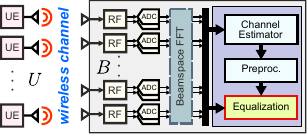}
	\caption{Overview of uplink processing in massive MU-MIMO. If the beamspace FFT is present, the subsequent operations are carried out in beamspace. In this paper we focus on the equalization implementation.}
	\label{fig:BD_system}
\end{figure}

We now describe an example application for the proposed VP number: beamspace processing in millimeter-wave massive multi-user (MU) multiple-input multiple-output (MIMO) systems.
Consider the system in \fref{fig:BD_system} and
let $\bms \in\setS^U $ be the vector of data symbols transmitted by all UEs, where $\setS$ is a discrete constellation set, e.g., 16-QAM, with the power constraint $\Ex{}{|s_u|^2} = E_s$, $u=1,\ldots,U$. The vector of baseband received signals at the basestation (BS) is given by 
\begin{align} \label{eq:AD_model}
	\bar{\bmy} = \bar{\bH}\bms + \bar{\bmn},
\end{align} 
where $\bar{\bH} \in \mathbb{C}^{B \times U}$ is the uplink channel matrix and $\bar{\bmn}$~is a complex Gaussian noise vector with per-entry variance~$N_0$. 
Equation \fref{eq:AD_model} is referred to as the \emph{antenna-domain} system model, as it models the signals received at the BS antennas. 

A prominent data detector commonly employed in massive MU-MIMO uplink processing is the linear minimum mean squared error (LMMSE) detector, which consists of (i) preprocessing, where $\bar{\bW} = (\bar{\bH}^H \bar{\bH} + N_0/E_s \bI_U)^{-1}\bar{\bH}^H$ is computed from the channel matrix ${\bar{\bH}}$, and (ii) equalization, where the estimate of the transmitted symbols is computed as $\hat{\bms} = \bar{\bW} \bar{\bmy}$. 

\subsubsection*{Beamspace Processing}

By applying a spatial DFT to the antenna-domain vector $\bar{\bmy}$ received at a uniform linear antenna array, we arrive at the following \emph{beamspace} system model:
\begin{align} \label{eq:BD_model}
	\bmy = \bF\bar{\bmy} = \bF\bar{\bH}\bms + \bF\bar{\bmn} 	= \bH \bms + \bmn.
\end{align} 
Here, $\bmy\in\complexset^B$ is the \emph{beamspace} receive vector,  $\bF\in\complexset^{B\times B}$ is the unitary DFT matrix, $\bH = \bF\bar{\bH}$ is the beamspace MIMO channel matrix, and $\bmn = \bF\bar{\bmn}$ is the beamspace-equivalent noise vector, which has the same statistics as $\bar{\bmn}$, since $\bF$ is unitary.
Therefore, the beamspace system model in \fref{eq:BD_model} is statistically equivalent to the antenna-domain system model, and data detection using both models gives the same result.

\begin{figure}[tp]
	\centering	
	\begin{subfigure}{0.48\columnwidth}
		\includegraphics[width=\columnwidth]{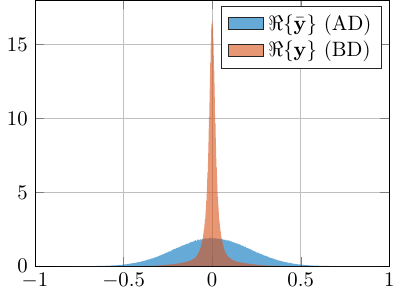}
		\caption{PDF of $\Re\{\bar{\mathbf{y}}\}$ and $\Re\{\mathbf{y}\}$}
		\label{fig:PDF_H64_LOS}
	\end{subfigure}
	\hfill
	\begin{subfigure}{0.5\columnwidth}
		\includegraphics[width=\columnwidth]{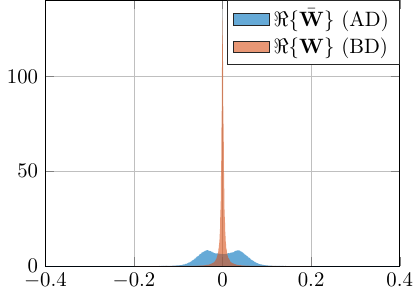}
		\caption{PDF of $\Re\{\bar{\mathbf{W}}\}$ and $\Re\{\mathbf{W}\}$}
		\label{fig:PDF_H64_NLOS}
	\end{subfigure}
	\caption{Empirical PDF of the real part of the entries of (a) $\bar{\bmy}$ and ${\bmy}$, and (b) $\bar{\bW}$ and ${\bW}$, using LoS channels generated by QuaDRiGa~\cite{QuaDRiGa_tech_rpt}.}
	\label{fig:PDF}
\end{figure}

Millimeter wave massive MIMO channel matrices $\bH$ are approximately sparse in beamspace, especially in line-of-sight (LoS) channels \cite{seyedhadi_thesis, lee16}. As a result, the beamspace received vectors and the LMMSE equalization matrices are also approximately sparse \cite{seyedhadi_thesis}. The sparsity of beamspace variables is illustrated by their spiky empirical PDFs in \fref{fig:PDF}.

Recently, beamspace sparsity has been exploited to reduce computational complexity of linear equalization in beamspace, i.e., $\hat{\bms} = \bW \bmy$, \cite{SeyedHadi20b, mirfarshbafan24, Mahdavi19}. The sparsity of beamspace variables results in higher dynamic range of beamspace signals, as demonstrated in the following section. In a fixed-point design, the higher dynamic range of signals calls for higher bitwidth, which in turn increases the silicon area and power consumption of the design. This aspect has been overlooked in literature with hardware implementation of beamspace processing. We will show how using VP can help mitigate this problem.

\subsection{Simulations} \label{sec:sims} 
To corroborate the observation that beamspace signals require larger bitwidth compared to antenna-domain signals, we performed the following experiment. For a massive MIMO BS with a uniform linear array of $B=64$ antennas communicating with $U=8$ single-antenna UEs with 16-QAM symbols, we generated $10^5$ antenna-domain uplink channel matrices $\bar{\bH}$ using the QuaDRiGa simulator \cite{QuaDRiGa_tech_rpt} in LoS conditions and one uplink received vector $\bar{\bmy}$ for each of these channel matrices at $20$\,dB SNR. For each channel matrix we also computed the LMMSE equalization matrix $\bar{\bW} = (\bar{\bH}^H \bar{\bH} +  N_0/E_s \bI_U)^{-1}\bar{\bH}^H$. Additionally, we computed the corresponding beamspace channel matrices $\bH = \bF \bar{\bH}$, the beamspace received vectors $\bmy = \bF \bar{\bmy}$ and the LMMSE matrix ${\bW} = (\bH^H \bH + N_0/E_s \bI_U)^{-1}\bH^H$. 
In order to unify the dynamic range of antenna-domain and beamspace variables, we scaled all instances of $\bar{\bW}$ with a single scalar such that the real and imaginary components of entries of all $\bar{\bW}$s lie in $(-1,1)$. We did a similar normalization for $\bW$, $\bar{\bmy}$, and ${\bmy}$, each with their own scalar.

\begin{figure}[t]
	\centering
	\includegraphics[width=0.65\linewidth]{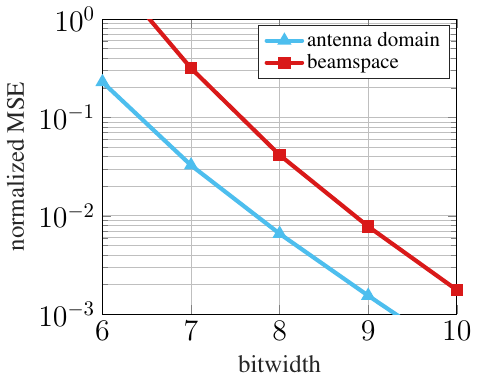}
	\caption{Normalized MSE vs the bitwidth of operands of the equalization operation in antenna-domain ($\hat{\bms} = \bar{\bW} \bar{\bmy}$) and beamspace ($\hat{\bms} = \bW \bmy$).}
	\label{fig:NMSE}
\end{figure}

For each of the $10^5$ pairs of $\bar{\bW}$ and $\bar{\bmy}$, we computed the unquantized matrix-vector product $\hat{\bms} = \bar{\bW}\bar{\bmy}$, as well as its quantized version $\hat{\bms}^{\text{q}}_W=f_{W,W-1}(\bar{\bW})f_{W,W-1}(\bar{\bmy})$, where $f_{W,F}(.)$ is the two's complement fixed-point quantization function with a total bitwidth of $W$ and $F$ fractional bits. Note that we chose $W-1$ fractional bits since we scaled the inputs such that the real and imaginary parts of all entries are between $-1$ and $1$, so we only need $1$ sign bit and the rest are fractional bits. Similarly, we computed the unquantized dot product with the beamspace inputs $\hat{\bms} = \bW \bmy$ and the quantized version $\hat{\bms}^{\text{q}}_W=f_{W,W-1}({\bW})f_{W,W-1}({\bmy})$. Note that for the quantized version, we only quantized the inputs and the multiplication itself was carried out with floating-point arithmetic, which means the only source of error in the quantized dot  product is the quantization of the inputs. We then computed the normalized mean squared error (NMSE) for $W = 6, \ldots, 10$, for both antenna-domain and beamspace dot products as 
\begin{align} \label{eq:NMSE}
	\text{NMSE}_W = 
	\frac{\Ex{}{\| \hat{\bms}^{\text{q}}_W - \hat{\bms} \|_2^2}}{ \Ex{}{\| \hat{\bms} \|_2^2}}.
\end{align} 

The results of this experiment are shown in \fref{fig:NMSE}. We observe that the quantized dot product using beamspace inputs requires around $1.2$ bits more than the quantized dot product using antenna-domain inputs to achieve the same NMSE. 
This is a result of the fact that the distribution of entries of the antenna-domain and beamspace inputs are significantly different, as illustrated in \fref{fig:PDF}; i.e., the majority of entries of beamspace inputs are concentrated around zero, while the antenna-domain inputs have a more spread distribution over the support set. 

\section{VLSI Implementation} \label{sec:vlsi}

As discussed in \fref{sec:beamspace}, the inputs of beamspace equalization ($\bW$ and $\bmy$) require larger bitwidth compared to the inputs of the antenna-domain equalization ($\bar{\bW}$ and $\bar{\bmy}$). In this section, we present VLSI architectures for both antenna-domain and beamspace equalization. For beamspace equalization, we present two variants: (i)~a purely fixed-point design and (ii)~a VP-based design in which the multiplications are carried out using VP inputs. Our goal is to demonstrate the effectiveness of the proposed VP format in reducing the area and power consumption of designs involving high-dynamic-range signals, compared to a corresponding FXP implementation.

\subsection{Matrix-Vector Multiplier (MVM) Architectures} \label{sec:archs}

\begin{figure*}[tp]
	\centering
	\begin{subfigure}[t]{0.28\textwidth}
		\includegraphics[width=\textwidth]{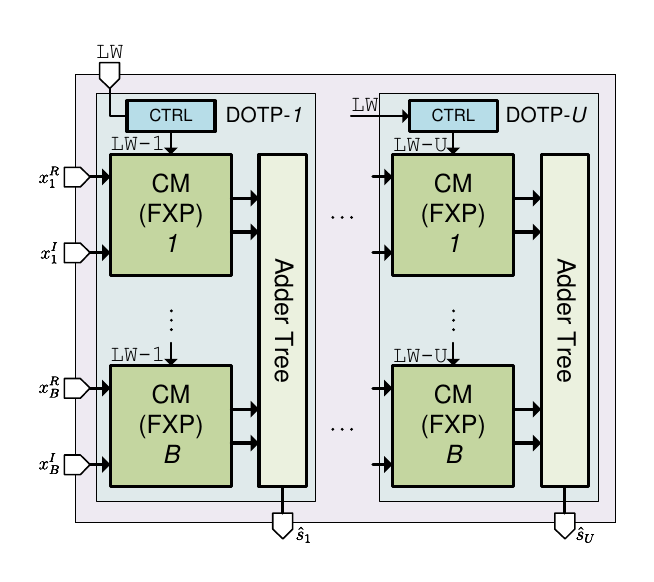}
		\caption{A-FXP}
		\label{fig:A-FXP}
	\end{subfigure}
	\hfill
	\begin{subfigure}[t]{0.315\textwidth}
		\includegraphics[width=\textwidth]{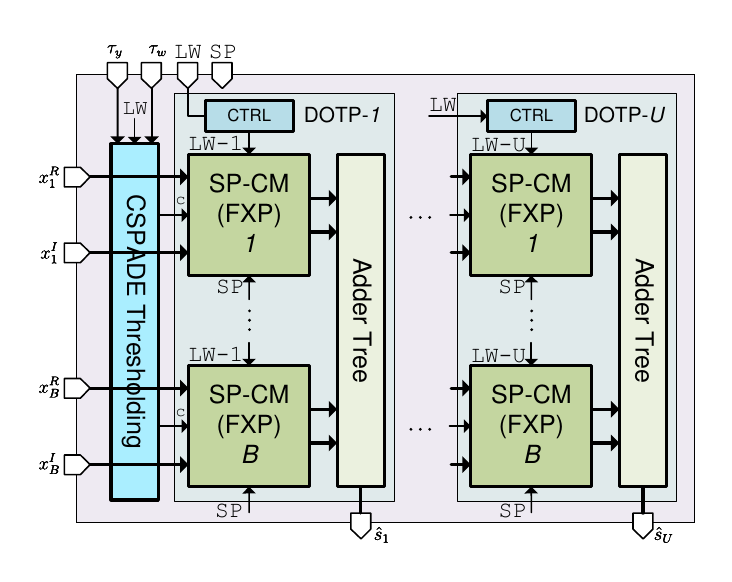}
		\caption{B-FXP}
		\label{fig:B-FXP}
	\end{subfigure}
	\hfill
	\begin{subfigure}[t]{0.37\textwidth}
		\includegraphics[width=\textwidth]{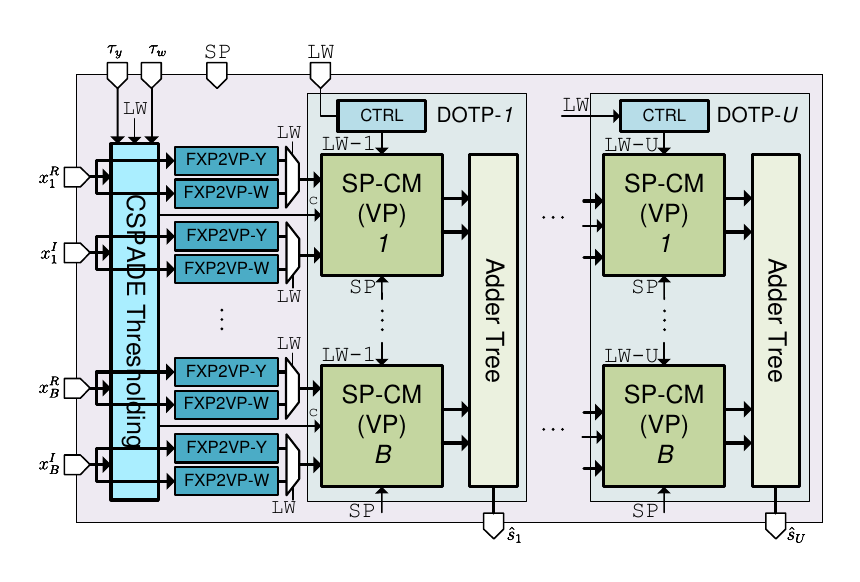}
		\caption{B-VP}
		\label{fig:B-VP}
	\end{subfigure}
	
	\caption{Overall architecture of A-FXP, B-FXP and B-VP equalizer designs.}
	\label{fig:archs}
\end{figure*}

\begin{figure}[t]
	\centering
	\includegraphics[width=0.7\linewidth]{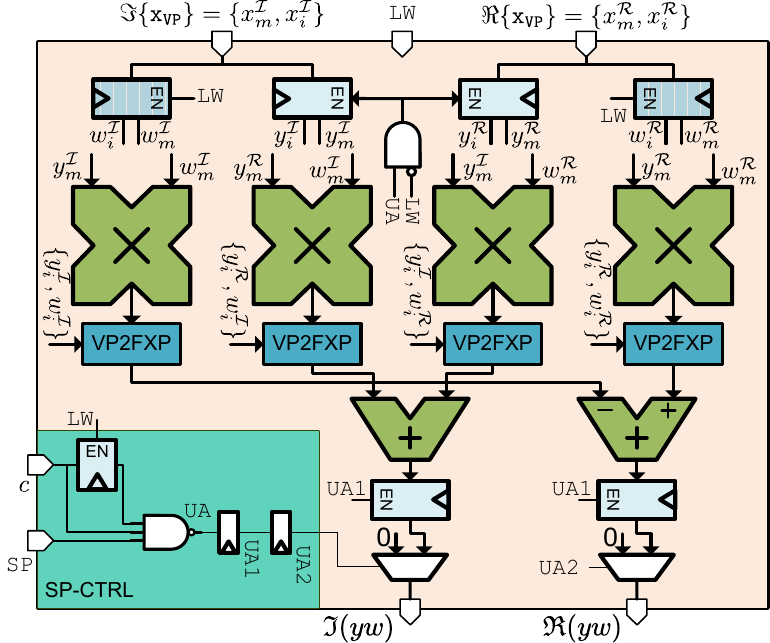}
	\caption{Internal architecture of SP-CM (VP).}
	\label{fig:SP_CM_VP}
\end{figure}

\fref{fig:archs} illustrates the three matrix-vector multiplier (MVM) variants considered in this paper: (a)~fixed-point MVM for antenna-domain equalization referred to as A-FXP, (b)~fixed-point CSPADE MVM~\cite{mirfarshbafan25} for beamspace equalization referred to as B-FXP, and (c)~VP-based CSPADE MVM for beamspace equalization referred to as B-VP.
The core component of all variants is the fully unrolled MVM illustrated in \fref{fig:A-FXP}, which consists of $U$ dot product units (abbreviated as DOTP), each consisting of $B$ complex-valued multipliers (CMs) and a $B$-operand internally pipelined adder tree to sum up the partial products. A load weight (\texttt{LW}) signal indicates whether the input ports $x_1$ to $x_B$ carry a row of the equalization matrix $\bar{\bW}$, or a received vector $\bar{\bmy}$. Once all rows of the equalization matrix are loaded in the respective dot product units, the MVM performs one equalization operation per clock cycle. 
The architectures in Figures~\ref{fig:B-FXP} and \ref{fig:B-VP} perform CSPADE-based equalization in order to reduce power consumption by exploiting the sparsity of $\bW$ and $\bmy$. Hence, in addition to the MVM core, they contain CSPADE thresholding circuitry, to skip  partial products for which the magnitude of both operands are below predetermined thresholds---this achieves significant dynamic power savings \cite{mirfarshbafan21b}. Furthermore, B-FXP and B-VP are made of CSPADE-enabled complex-valued multipliers (SP-CMs), which are slightly different than the CMs used in A-FXP. The hardware overhead due to the CSPADE operation in B-FXP and B-VP is negligible.
The difference between B-FXP and B-VP is that (i)~the B-VP contains a pair of FXP2VP converters for each real and imaginary FXP input, one for converting the $\bmy$ inputs and one for the $\bW$ inputs coming from the same ports, and (ii)~the SP-CMs used in B-VP perform VP multiplication and convert the result back to FXP using VP2FXP converters after each real-valued multiplier (RM). The advantage of the B-VP architecture is that the RMs inside its CMs are smaller thanks to the VP-based multiplication, which reduces the bitwidth of the multiplier operands (cf.~\fref{sec:par}).

\subsection{Complex-Valued Multipliers (CMs)} \label{sec:CMs}
The internal architecture of the SP-CM (VP) is depicted in \fref{fig:SP_CM_VP}.
The SP-CM (FXP) has the same architecture except that it does not contain the VP2FXP converters, as all signals are in FXP format. The CMs used in A-FXP also have the same architecture except that they do not contain the VP2FXP converters nor the CSPADE controller that generates conditional muting signals for CSPADE operation.

\subsection{Parameter Optimization} \label{sec:par}
In order to enable a fair assessment  of the potential of VP in representing high-dynamic-range signals with smaller significand bitwidth, for each of the three design variants, we fully optimized the fixed-point parameters (i.e., the total bitwidth and the number of fractional bits) and the VP parameters (i.e., the significand bitwidth and the exponent list) so that the bit error rate (BER) of the LMMSE equalization produced by the proposed architecture with the optimized parameters does not show a visible gap to the BER of floating-point LMMSE equalization.
The optimized parameters are listed in \fref{tbl:params}. In this table, $(W,F)$ designates a $W$-bit fixed-point number with $F$ fractional bits, and $(M, \bmf)$ designates a VP number with an $M$-bit significand and the exponent list given by $\bmf$. These parameters confirm the observation from our NMSE simulations in \fref{sec:sims} that beamspace inputs need approcimately $1$-to-$2$ bits more than the antenna-domain counterparts to achieve similar accuracy.

\begin{table}[tp]
	\centering
	{\caption{FXP and VP parameters of equalization inputs}
		\label{tbl:params} 
		\renewcommand{\arraystretch}{1.1}
		\begin{minipage}[c]{1\columnwidth}
			\centering
			
			\begin{tabular}{@{}cc|cc|cc@{}}
				\toprule
				\multicolumn{2}{c|}{\,A-FXP} & \multicolumn{2}{c}{\,B-FXP} & \multicolumn{2}{|c}{\,B-VP} \\
				\midrule
				$\bar{\bmy}$ & $(7,1)$ & $\bmy$ & $(9,1)$ & $\bmy$ & $(7,[1,-1])$\\				
				$\bar{\bW}$ & $(11,10)$ & $\bW$ & $(12,11)$ & $\bW$ & $(7, [11,9,7,6])$\\
				\bottomrule
			\end{tabular}
			
	\end{minipage}}
\end{table} 

\section{Implementation Results} \label{sec:impl}

We now present post-layout implementation results for a 22\,nm FDSOI process for the three equalizer architectures from \fref{sec:vlsi}, and for a massive MIMO system with $B=64$ and $U=8$. In all three designs, we used the same timing and area constraints, and the resulting implementations achieved similar slacks, enabling a direct comparison of area and throughput.
\subsection{Area and Power}

\begin{figure*}[tp]
	\centering
	\begin{subfigure}[t]{0.31\textwidth}
		\includegraphics[width=\textwidth]{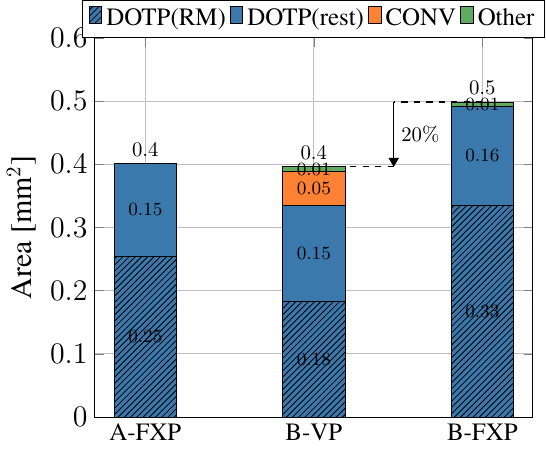}
		\caption{Area}
		\label{fig:impl_area}
	\end{subfigure}
	\hfill
	\begin{subfigure}[t]{0.32\textwidth}
		\includegraphics[width=\textwidth]{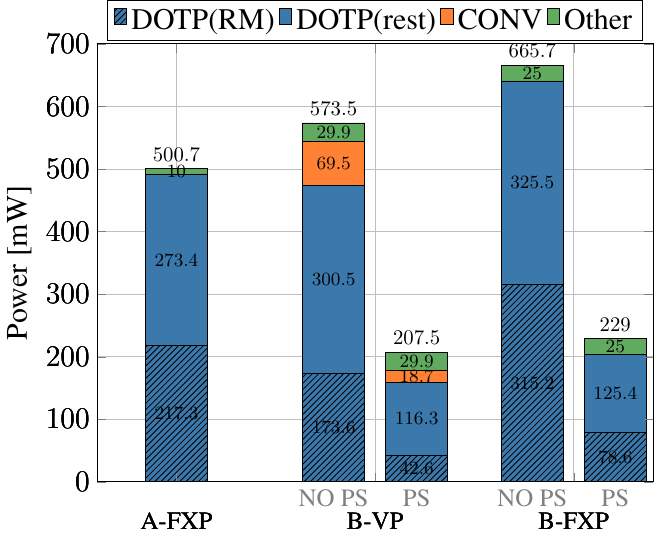}
		\caption{Power with LoS stimuli} 
		\label{fig:pwr_LOS}
	\end{subfigure}
	\hfill
	\begin{subfigure}[t]{0.32\textwidth}
		\includegraphics[width=\textwidth]{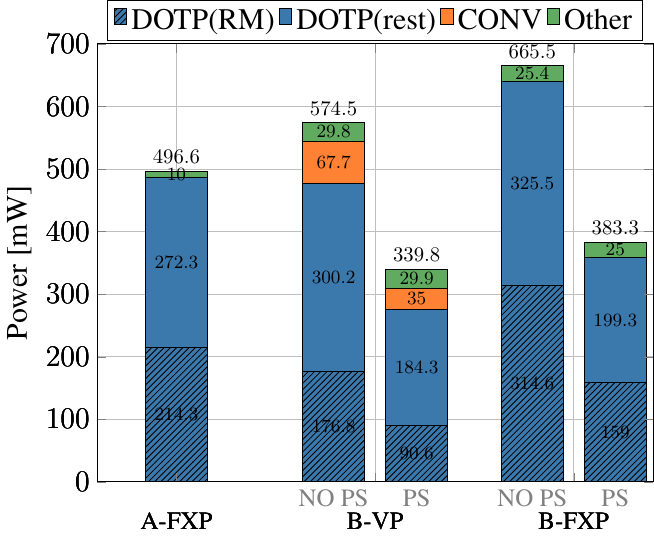}
		\caption{Power with non-LoS stimuli} 
		\label{fig:pwr_NLOS}
	\end{subfigure}

	\caption{Area (a) and power breakdown of A-FXP, B-VP, and B-FXP implementations with LoS and non-LoS stimuli (b,c). In (b) and (c), we show the results with power savings (PS) and without power savings (NO PS) activated for the B-VP and B-FXP designs.}
	\label{fig:impl}
\end{figure*}

\fref{fig:impl_area} shows the area breakdown. The blue part of each bar shows the total area of the DOTP units, with the hatched part indicating the area due to the RMs. The B-VP design contains FXP2VP converters at the inputs of the DOTP units and VP2FXP converters after each RM; the aggregate area of these converters is shown in the orange part (CONV). The B-VP and B-FXP designs additionally contain CSPADE thresholding circuitry and some multiplexers, whose area is lumped into `Other' in \fref{fig:impl_area}. 

We observe that the larger bitwidth of beamspace signals results in $25\%$ larger area in B-FXP compared to A-FXP.
Furthermore, the area of the RMs, which constitute $66\%$ of the total area of B-FXP, reduces from $0.33\,\text{mm}^2$ to $0.18\,\text{mm}^2$ in B-VP thanks to the VP format, which results in $20\%$ overall area savings compared to B-FXP, despite the converter overhead.

Figures~\ref{fig:pwr_LOS} and \ref{fig:pwr_NLOS} show the power consumption breakdown of the three designs, using post-layout simulations with node switching rates extracted from stimuli with LoS and non-LoS channels, respectively. All three designs are running at a clock frequency of $1$\,GHz. From this figure, we see that the power savings achieved through the VP format is less than the area savings. The main reason is that while RMs occupy $66\%$ of the total area of the B-FXP design, they stand for only $34\%$ to $47\%$ of the B-FXP power (depending on the stimuli and whether CSPADE power savings is activated). Noting that the VP-based design only affects the RMs, the power savings achieved in VP-based design is smaller than the area savings.
Nonetheless, as we see in \fref{fig:impl}, the B-VP design consumes $10\%$ to $14\%$ less power than the B-FXP design, with and without CSPADE power savings enabled, respectively.

\subsection{Comparison with Custom FLP} \label{sec:custom_FLP}

In order to demonstrate the advantages of the proposed VP format compared to a fully customized floating-point format, we performed the following experiment. We designed an array of $U=8$ CSPADE-based complex-valued multiply-accumulate (CMACs), which performs the same equalization operation as the circuit in \fref{fig:B-FXP} over $B$ clock cycles. We implemented two versions of this CSPADE CMAC array: (i)~one using a unified fully customized FLP for all signals and (ii)~another one using the VP format for the inputs of multipliers (additions and subtractions are done in FXP). In order to minimize the area of the custom FLP design, we minimized the mantissa bitwidth and number of exponent bits such that the FLP design achieves the same performance as the VP design with the parameters given in  \fref{tbl:params}. Through extensive simulations we found that the optimal FLP format consists of one sign bit, a $9$-bit mantissa, and a $4$-bit exponent.

For the FLP design, we configured the floating-point arithmetic components not to implement the IEEE-compliant features; i.e., no support for not-a-numbers (NaNs) and denormal numbers. IEEE-compliant arithmetic components are around $1.5 \times$ larger than the non-compliant variants for the same timing constraints. With all these optimizations, the area of the FLP-based CMAC array is $3.4 \times$ larger than the area of the VP-based design and consumes on average $3 \times$ more power than the VP-based design. This result, combined with the results shown in \fref{fig:impl_area}, confirm the effectiveness of the proposed VP format, as it achieves $70\%$ and $20\%$ area savings compared to the optimized FLP and FXP designs, respectively.

\section{Conclusion}

We have proposed a novel number format, called variable-point (VP). In VP, the exponent bits do not encode a contiguous exponent range---instead, they serve as an index for a user-defined exponent list. This approach enables non-uniform exponent spacing optimized for the target application. As a result, VP can represent high-dynamic-range signals using a lower-resolution significand than a fixed-point format, resulting in (often significant)  area and power savings.

As a case study, we have implemented MVM engines for equalization in millimeter-wave massive MU-MIMO in antenna domain and beamspace. Equalization in beamspace involves high-dynamic-range signals, resulting in larger FXP implementation than the antenna-domain counterpart. Through post-layout implementation results on a 22nm CMOS process, we demonstrated that the VP-based MVM design offers $20\%$ area reduction compared to a fully optimized fixed-point implementation. Furthermore, we showed that a VP-based MAC array is $3.4\times$ smaller than a fully customized floating-point MAC array.

While we have only studied the efficacy of VP arithmetic for a communications application, we are convinced that VP numbers can also improve the efficiency of customized circuits for machine learning accelerators.

\end{document}